\documentclass[12pt,preprint]{aastex}
\shorttitle{Large scale structure in the CDFS}
\shortauthors{Gilli et al.}

\def\chandra{{\it Chandra }}

\def\lum{\rm\;erg\;s^{-1}}

\begin{document}


\title{Tracing the large scale structure in the \chandra Deep Field South}


\author 
{
%CDFS + K20 collaboration
R. Gilli\altaffilmark{1,2}, 
A. Cimatti\altaffilmark{1}, 
E. Daddi\altaffilmark{3},
G. Hasinger\altaffilmark{4}, 
P. Rosati\altaffilmark{3},
G. Szokoly\altaffilmark{4}, 
P. Tozzi\altaffilmark{5}, 
J. Bergeron\altaffilmark{3}, 
S. Borgani\altaffilmark{6}, 
R. Giacconi\altaffilmark{2,7},
L. Kewley\altaffilmark{8},
V. Mainieri\altaffilmark{3,9},
M. Mignoli\altaffilmark{10},
M. Nonino\altaffilmark{5}, 
C. Norman\altaffilmark{2,11}, 
J. Wang\altaffilmark{2}, 
G. Zamorani\altaffilmark{10},
W. Zheng\altaffilmark{2}, 
and A. Zirm\altaffilmark{2} 
%Roberto Gilmozzi\altaffilmark{2}, 
%Norman Grogin\altaffilmark{2}, 
%Anton Koekemoer\altaffilmark{2}, 
%and Ethan Schreier\altaffilmark{2}
}

\altaffiltext{1}{Istituto Nazionale di Astrofisica (INAF) -
Osservatorio Astrofisico di Arcetri, Largo E. Fermi 5, 50125 Firenze, Italy}

\altaffiltext{2}{The Johns Hopkins University, Homewood Campus,
Baltimore, MD 21218}

\altaffiltext{3}{European Southern Observatory,
Karl-Schwarzschild-Strasse 2, Garching, D-85748, Germany}

\altaffiltext{4}{Max-Planck-Institut f\"ur extraterrestrische Physik, 
Postfach 1312, D-85741 Garching, Germany}

\altaffiltext{5}{Istituto Nazionale di Astrofisica (INAF) -
Osservatorio Astronomico, Via G. Tiepolo 11, 34131 Trieste, Italy}

\altaffiltext{6}{INFN, c/o Dip. di Astronomia dell'Universit\`a,
Via G. Tiepolo 11, 34131 Trieste, Italy}

\altaffiltext{7}{Associated Universities, Inc. 1400 16th Street, NW,
Suite 730, Washington, DC 20036}

\altaffiltext{8}{Harvard-Smithsonian Center for Astrophysics, 60
Garden Street, Cambridge, MA 02138}

\altaffiltext{9}{Dip. di Fisica, Universit\`a degli Studi Roma Tre,
Via della Vasca Navale 84, I-00146 Roma, Italy}

\altaffiltext{10}{Istituto Nazionale di Astrofisica (INAF) -
Osservatorio Astronomico di Bologna, Via Ranzani 1, 40127 Bologna, Italy}

\altaffiltext{11}{Space Telescope Science Institute, 3700 San Martin
Drive, Baltimore, MD 21218}

%\altaffiltext{4}{Astrophysikalisches Institute Potsdam, An der
%Sternwarte 16, Potsdam, D-14482, Germany}

\begin{abstract}

We report the discovery of large scale structures of X-ray sources in
the 1Msec observation of the \chandra Deep Field South. Two main
structures appear as narrow ($\Delta z \lesssim 0.02$) spikes in the
source redshift distribution at z=0.67 and z=0.73, respectively. Their
angular distribution spans a region at least $\sim 17 $ arcmin wide,
corresponding to a physical size of $7.3 h_{70}^{-1}$ Mpc at a
redshift of $z\sim 0.7$ ($\Omega_m=0.3$, $\Omega_{\Lambda}=0.7$).
These spikes are populated by 19 sources each, which are mainly
identified as Active Galactic Nuclei (AGN). Two sources in each spike are
extended in X-rays, corresponding to galaxy groups/clusters embedded
in larger structures. The X-ray source redshift distribution shows
other spikes, the most remarkable at z=1.04, 1.62 and 2.57. This is
one of the first evidences for large scale structure traced by X-ray
sources and for spatial clustering of X-ray selected AGN. The X-ray
data have been complemented with the spectroscopic data from the K20
near infrared survey \citep{cima02}, which covers $\sim 1/10$ of the
X-ray field. Also in this survey the source redshift distribution
shows several spikes. Two narrow structures at z=0.67 and z=0.73
(again with $\Delta z \sim 0.02$) are the most significant features,
containing 24 and 47 galaxies, respectively. While the K20 structure
at z=0.73 is dominated by a standard galaxy cluster with a significant
concentration around a central cD galaxy and morphological
segregation, the galaxies at z=0.67 constitute a loose structure
rather uniformly distributed along the K20 field. 

Moreover, we find a very good correlation (almost one-to-one)
also between less prominent peaks detected in the redshift
distributions of X-ray and K20 sources. In particular, at $z<1.3$ we
find that 5 out of the 6 more significant K20 peaks have a
corresponding peak in the X-ray selected sources and, similarly, all
the 5 X-ray peaks below that redshift have a corresponding K20
peak. Since the K20 survey sensitivity drops beyond $z\sim1.3$,
structures at higher redshift are traced only by the X-ray sources.
This correlation suggests that AGN (from the X-ray data) and
(early-type) galaxies (from the K20 survey) are tracing the same
underlying structures.

We also compared the X-ray and K-band catalogs to search for enhanced
X-ray activity in the sources in the two main redshift spikes. While
in the structure at z=0.73 the fraction of X-ray sources is the same
as in the field, in the structure at z=0.67 it is higher by a factor
of $\sim 2$, suggesting that X-ray activity may be triggered
preferentially in the structure at z=0.67. Given the limited
statistics, this result is significant only at $\sim 2\sigma$ level.

\end{abstract}

\section{Introduction} \label{introduction}

While the large scale structure of the Universe at $z<1$ is usually
mapped through galaxy surveys, AGN surveys are a powerful tool to
study the clustering of high redshift objects. AGN clustering has been
extensively studied and detected at optical wavelengths
\citep{shank87, lafra98, croom01}, where objects are mainly selected
by means of their strong UV excess and include almost exclusively
unobscured-type 1 AGN.
%
%The most recent work is based on a sample of $> 10^4$ QSO taken from
%the 2dF QSO Redshift Survey (2QZ, Croom et al. 2001), which measured
%the QSO clustering in the broad redshift range $z=0.3-2.9$, confirming
%previous results and showing that the QSO clustering is comparable to
%that of local ($z \sim 0.05$) optically selected galaxies (Tucker et
%al. 1997; Ratcliffe et al. 1998).
%Georgantopoulos \& Shanks (1994) analyzed the spatial clustering of a
%sample of 200 Seyfert galaxies observed with IRAS, selected by
%their warm infrared colors. 
%Although  selection bias in favor of unobscured
%AGN is indeed present also in this sample, since warmer colors are
%more frequently found in Seyfert 1s than in Seyfert 2s (see e.g.
%Risaliti et al. 2000). 
%
An advantage of the X-ray selection, especially in the hard X-rays,
resides in the capability of detecting also obscured AGN, which, based on
population synthesis models for the X-ray background, are believed to
be a factor of 4--10 more abundant than unobscured AGN \citep{comas95,
gilli01}.

While angular clustering of X-ray selected AGN was detected by several
authors \citep{vikh95, akylas00, giac01}, yet there are no direct
measurements of their spatial clustering. \citet{boyle93} studied
the AGN at $z<0.2$ in the Einstein Medium Sensitivity Survey (EMSS,
Stocke et al. 1991), without finding any positive clustering
signal. \citet{carrera98} considered the AGN in the ROSAT
International X-ray Optical Survey (RIXOS, Mason et al. 2000) and in
the Deep ROSAT Survey \citep{boyle94}, detecting only a weak
($\sim 2\sigma$) clustering signal on scales $<60-120\:h_{70}^{-1}$ Mpc for
the RIXOS AGN subsample in the redshift range $z=0.5-1.0$.

%Significant clustering
%signal was instead detected from angular (i.e. 2D) correlations by
%several authors: Akylas et al. (2000) based on the ROSAT All Sky
%Survey (RASS, Voges et al. 1995); Vikhlinin \& Forman (1995) from a
%compilation of ROSAT PSPC deep pointings and finally Giacconi et
%al. (2001) from the first 130 ksec observation of the \chandra Deep
%Field South (Rosati et al. 2001).

The \chandra Msec surveys in the Deep Field South (CDFS, Rosati et al.
2002) and North (CDFN, Brandt et al. 2001) will allow a step forward
in the detection of X-ray source structures. A large spectroscopic
identification program down to very faint magnitudes ($R<25.5$) is
underway in the CDFS, which is expected to provide reliable redshift
estimates for $\sim 70\%$ of the X-ray sample. Although $10-20\%$ of
the sources in the deep X-ray surveys are identified as
normal/starburst galaxies, still AGN are the dominant population
\citep{giac01, barger02}. In this paper we will consider the redshift
distribution of those CDFS sources for which a robust redshift
determination is already available ($\sim 38\%$ of the full
sample). This distribution shows two prominent spikes at $z=0.67$ and
$z=0.73$ (see also Hasinger 2002) as well as several other spikes
at higher redshift. Narrow spikes are commonly detected in the
redshift distribution of optical galaxies observed in deep pencil beam
surveys \citep{lefev96, cohen99}, and are now being observed also
among X-ray selected objects.  Narrow structures at $z\sim 0.8$ and
$z\sim 1$, similar to those observed in the CDFS, are indeed observed
also in the CDFN \citep{barger02}.

In this paper the features in the redshift distribution of X-ray
sources in the CDFS are compared with those observed in the K20
survey \citep{cima02}, which efficiently selects massive galaxies
in a broad redshift range and covers part of the CDFS field. In
addition, the X-ray and K20 catalogs are cross correlated to search
for any enhanced X-ray activity in the two main redshift structures
with respect to the field. We note that a concentration of
objects at z=0.67 in the CDFS (mostly early type galaxies) was also
reported by \citet{cwg01}, who selected objects with $K\leq 19.5$ and
$J-K$ colors redder than the stellar sequence.

The analysis of the spatial clustering of the full X-ray sample
and the comparison with the spatial clustering of K20 sources
will be presented in a future paper.

Throughout this paper we will use a cosmology with $H_0=70$ km
s$^{-1}$ Mpc$^{-1}$, $\Omega_m=0.3$, $\Omega_{\Lambda}=0.7$.

\section{X-ray sources} \label{xray}

The CDFS has been observed for 1Msec with ACIS-I, and represents one
of the deepest X-ray surveys to date \citep{rosati02}. Limiting fluxes
of $5.5\;10^{-17}$ erg cm$^{-2}$ s$^{-1}$ and
$4.5\;10^{-16}$ erg cm$^{-2}$ s$^{-1}$ have been reached in the 
0.5-2 keV and 2-10 keV band, respectively; 
346 sources have been detected in the whole 0.1
deg$^2$ field. The full X-ray catalog and the details of the detection
process have been presented by \citet{giac02}. The optical follow-up
was primarily performed using the FORS1 camera at the VLT. The
combined R band data cover a $13.6 \times 13.6$ arcmin field to
limiting magnitudes between 26 and 26.7. In the CDFS area not covered
by FORS mosaics, we used shallower data from the ESO Imaging Survey
(EIS, Arnouts et al. 2001). The optical identification process is
described in \citet{tozzi01} and \citet{giac01}. Optical spectroscopy
for most of the X-ray counterparts with R$<25.5$ has been obtained with
FORS1 during several observational runs at VLT. The analysis of
the spectroscopic data is almost completed. The details of the data
reduction, as well as the final redshift list, will be presented in
Szokoly et al. (in preparation). So far 169 redshifts have been
obtained. Quality flags have been assigned to the spectra, according
to their reliability. Here we consider only the 121 sources with
a quality flag $Q=2$ (excluding a few stars), where two or more
emission lines have been observed and the redshift determination is
unambiguous, plus 10 sources whose redshift is determined from the
cross-correlation with K20 sources. The sample includes sources
detected either in the soft (0.5-2 keV) or in the hard (2-10 keV) band
or both. The $1\sigma$ errors on the redshift
measurements are typically of the order of $\Delta\:z \sim 0.002$. The
redshift distribution is dominated by two large concentrations of
sources in the ranges $0.664 \leq z \leq 0.685$ and $0.725 \leq z \leq
0.742$ containing 19 objects each.
%
%We defined the width of the two structures from the redshift distribution
%of K20 sources, where the statistics is higher. Sources with
%$0.662<z<0.682$ are considered to be part of the first structure and will be
%referred in the following as {\it sources at $z=0.67$}. Sources with
%$0.724<z<0.744$ are considered to be part of the second structure, and will
%be referred to in the following as {\it sources at $z=0.73$}.
%Refining the choice of the redshift space covered by the two structures
%will not alter our results significantly. There is only one X-ray
%source and no K20 sources with measured redshift between these two
%stripes.
%
In the following we will refer to these two structures as {\it sources
at $z=0.67$} and {\it sources at $z=0.73$}, respectively.  These
striking structures can be seen in the upper panel of Figure~1, where
we plot the redshift distribution of X-ray sources in redshift bins of
$\Delta z=0.02$. Due to the fixed bin width, both structures appear to
contain 18 rather than 19 sources. Other possible, less prominent
structures (see Section~4) are clearly visible in this plot.

As shown in Figure~2, sources at z=0.67 and z=0.73 are mostly
distributed in one half of the field. Since the spectroscopic coverage
of the CDFS is rather uniform (Szokoly et al., in preparation), the
higher concentration of sources in half of the field is not likely to
be an artifact. The structures extend to a scale of $\sim 17$ arcmin,
corresponding to a linear physical size of 7.3 Mpc at $z\sim
0.7$. This is likely to be a lower limit, since the source angular
distribution appears to be limited only by the finite size of the
X-ray image. Two extended sources (CDFS 560, 645), identified as two
galaxy groups, are present in the structure at z=0.67. Similarly, two
extended sources (CDFS 594, 566) belong to the structure at z=0.73:
one is identified as a galaxy group, the other is associated with the
galaxy cluster in the K20 survey (see next Section and Fig. 2 and 3).
The full list of CDFS extended sources is given in \citet{giac02}.
Preliminary results on the CDFS extended sources are being obtained
from XMM observations: while CDFS 566 is not sufficiently extended to
be resolved by XMM, CDFS 594 appears to be a factor of $\sim 4$ more
extended than measured by \chandra (J. Bergeron, priv. comm.).

In addition to the above discussed structures, smaller peaks are also
recognized in the X-ray source redshift distribution. To assess the
significance of these structures we follow a procedure similar to that
adopted by \citet{cohen99}, who observed features in their galaxy 
redshift distribution with typical velocity dispersion of $\sim 300$
km/s. Sources are distributed in $V=c\;ln(1+z)$ rather than in
redshift, since $dV$ corresponds to local velocity variations granted
the Hubble expansion. The observed distribution is then smoothed with
a Gaussian with $\sigma_S = 300$ km/s to obtain the ``signal''
distribution. Since there is no a priori knowledge of the
``background'' field distribution, we heavily smoothed the observed
distribution with a Gaussian with $\sigma_B=1.5\: 10^4$ km/s and
considered this as the background distribution.

We then searched for possible redshift peaks in the signal
distribution, computing for each of them a signal-to-noise ratio
defined as SNR=$(S-B)/\sqrt{B}$, where S is the number of sources in a
velocity interval of fixed width $\Delta V=2000$ km/s around the
center of each peak candidate and B is the number of background
sources in the same interval. The value of $\Delta V$ was chosen to
optimize the SNR values for peaks populated by 3-5 sources. For more
populated peaks the quantity S may underestimate the actual number of
sources in each peak. Adopting the thresholds $S\geq 3$ and $SNR>3.8$
we find 7 peaks in the signal distribution. In order to estimate the
expected fraction of possibly ``spurious'' peaks arising from
background fluctuations, we simulated 1000 samples of 131 redshifts
randomly extracted from the smoothed background distribution and
applied our peak detection method to each simulated sample. The result
is that, with the adopted thresholds, the average number of spurious
peaks due to background fluctuations is 0.47. In only 13 of the simulated
samples 3 spurious peaks are detected, while in none of simulated samples
4 or more spurious peaks are detected.

The seven X-ray source peaks detected by our procedure with $S\geq 3$
and $SNR>3.8$ are listed in Table~1, where for each peak we give the
average redshift, the number of objects (N) in the peak and the
Poissonian probability of observing N sources given the background
value in the same velocity intervals where the N sources are found.

We checked our detection method with larger values of the smoothing
length for the signal distribution. We verified that, when increasing
$\sigma_S$, several peaks are lost; as an example, when the signal
smoothing is as large as 1500 km/s, corresponding to the velocity
dispersion of rich clusters, only the three most significant peaks at
z=0.67, 0.73 and 1.62 are detected. Therefore we will consider only
the results obtained with $\sigma_S =300$ km/s in the following.

\section{Sources from the K20 survey} \label{k20}

Part of the CDFS field has been covered by the K20 survey (see Cimatti
et al. 2002 and references therein), which includes all the sources
detected in the EIS K-band images to $Ks<20$. The K20 field (6.7 by
4.8 arcmin) covers $\sim 1/10$ of the X-ray field, and is shown as a
rectangle in Figure 2: 348 sources with $Ks<20$ have been found in this
area. Spectroscopic identification of the K20 sources has been
performed with FORS1 and FORS2 at the VLT. We will consider here only
the 258 galaxies with highest quality spectra and unambiguous redshift
determination. The typical error on the redshift estimate is
$\Delta\:z=0.002$. The redshift distribution of the K20 sources in
bins of $\Delta\:z=0.02$ is shown in the lower panel of Figure~1. In
strict analogy with the X-ray results, the redshift distribution of
the K20 galaxies has two prominent peaks at z=0.67 and z=0.73.

While in the X-rays the two structures at z=0.67 and z=0.73 appear
equally populated, in the smaller K20 field sources at 0.73 (47
objects) are a factor of $\sim 2$ more numerous than sources at 0.67
(24 objects). The redshift ranges defining the two spikes are the same
adopted for X-ray sources ($0.664 \leq z \leq 0.685$ and $0.725 \leq z
\leq 0.742$). Again, due to the fixed bin size adopted in Fig.~1, only
45 and 22 sources appear in the spikes at z=0.73 and z=0.67,
respectively.

Inspection of the distribution on the sky of the K20 sources together
with their spectral properties reveals that the two structures are
indeed very different. The structure at $z=0.73$ appears to be
dominated by a standard galaxy cluster, showing a central cD galaxy
and significant spectral segregation, with early type objects
concentrated around the cD galaxy. Overall, 25 out of 47 sources at
$z=0.73$ are classified as early type galaxies, 17 are emission line
galaxies, 4 have intermediate spectra, and 1 object has been
recognized as an AGN. Sixteen out of 19 K20 sources within a circle of
radius 1 arcmin ($\sim 440$ Kpc) around the central cD are early type
objects (see Fig.~3). On the contrary, sources at $z=0.67$ constitute
a loose structure with early and late type galaxies uniformly
distributed across the K20 field. For completeness, we note that 11 out
of 24 sources at $z=0.67$ are classified as early type galaxies, 11
are late type galaxies and 2 have intermediate spectra. 

We have searched for other peaks in the K20 redshift distribution
using the same procedure described in the previous Section. Adopting
the same parameters as in the analysis of the X-ray data for the
smoothing lengths ($\sigma_S=300$ km/s and $\sigma_B=1.5 \: 10^4$
km/s) and similar values for the thresholds in SNR and S ($SNR > 4.0$
and $S \geq 4$) we find 8 peaks. Also in this case the average number
of spurious peaks, obtained from 1000 simulated samples, is low
(0.11). None of the simulated samples contains 3 or more spurious
peaks. The central redshifts, number of sources, and Poissonian
probability related to the K20 peaks are quoted in Table~1.

%We also considered the effects of assuming a larger smoothing for the
%background distribution with $\sigma_B = 1.5 \: 10^5$ km/s.  This
%smoothing does not appear to be appropriate for the K20 distribution,
%since it overestimates the number of sources above z=1.3, where the
%signal distribution cut off, and, conversely, underestimates the
%number of sources below that redshifts, artificially increasing
%the number of detected peaks. Furthermore, such a background
%distribution cannot be used to extract the simulated samples, since it
%is not representative of the actual signal distribution. Then, we will
%concentrate only on the results obtained with $\sigma_S=300$ km/s and
%$\sigma_B=1.5 \: 10^4$ km/s quoted in Table~1. 
Inspection of Table~1 reveals that, besides the two prominent spikes
at z=0.67 and z=0.73, other peaks are in common between the K20 and
the X-ray data. The third most populated K20 peak (15 sources at
z=1.036) is detected also in the X-rays, being the third populated
peak of the X-ray redshift distribution (6 sources). Another two K20
peaks (at $z\sim$ 0.077 and 1.220) have a corresponding peak in the
X-ray source redshift distribution. Overall, 5 out of 8 peaks in the
K20 source redshift distribution have an X-ray peak counterpart. If we
restrict our analysis to the more significant K20 peaks with
Poissonian probability smaller than $10^{-3}$ and $5\: 10^{-4}$, the
fraction with X-ray counterpart is 5/6 and 4/4, respectively. The
fact that the less significant K20 peaks do not have a corresponding
X-ray peak may be expected since the total number of X-ray sources is
much lower (a factor of $\sim 2$) than that of K20 sources and is
spread over a wider redshift range.  Only about 30\% of the X-ray
sources in these peaks is within the area covered by the smaller K20
survey.  This means that, in general, the structures seen in the K20
survey are traced on wider scales by the X-ray sources. We note that
the high redshift peaks detected in the X-rays are not detected in the
K20 observations, whose sensitivity drops dramatically above $z\sim
1.3$.

%\begin{onecolumnfigure}
%{\epsfxsize=8.8cm\epsffile{kx_spikes}}
%\figcaption{}
%\end{onecolumnfigure}

%\begin{onecolumnfigure}
%{\epsfxsize=7.9cm\epsffile{ddbox}}
%\figcaption{}
%\end{onecolumnfigure}

\section {The X-ray source fraction in the redshift structures}

\subsection {Total X-ray sample}

Since enhanced nuclear or star forming activity (both marked by X-ray
emission) are likely to be produced by galaxy interactions in large
scale structures, we searched for any variation in the
X-ray to K20 source number ratio inside and outside the two structures
at z=0.67 and z=0.73. 

%We defined the width of the two structures from the redshift distribution
%of K20 sources, where the statistics is higher. Sources with
%$0.662<z<0.682$ are considered to be part of the first structure and will be
%referred in the following as {\it sources at $z=0.67$}. Sources with
%$0.724<z<0.744$ are considered to be part of the second structure, and will
%be referred to in the following as {\it sources at $z=0.73$}.
%Refining the choice of the redshift space covered by the two structures
%will not alter our results significantly. There is only one X-ray
%source and no K20 sources with measured redshift between these two
%stripes.

First, we simply divided the number of X-ray sources by the number of
K20 sources inside and outside the two structures to search for
variations with respect to the field. Extended X-ray sources (two in
each structure) have been excluded from this analysis. Despite the
significant spectroscopic incompleteness in the X-ray sample ($\sim
60\%$ of the CDFS sources have no secure redshift determination yet), the
results of this analysis should not be biased as long as the X-ray
spectroscopic incompleteness is ``random'' with respect to the
probability of an X-ray source being inside or outside an
overdensity. Since the redshift distributions of the two samples are
different, in estimating the number of sources in the ``field'' we
considered only the sources not belonging to the two structures in a
redshift range centered at the redshift of the two
structures. Unfortunately, given the limited statistics, this range
can not be too small. Adopting the redshift range 0.4--1.0, we find
that the X-ray to K-band number ratio is $0.33\pm 0.07$ in the field
(this ratio would be similar, but with a larger uncertainty, if we
adopted the narrower redshift range 0.5--0.9), $0.36\pm 0.10$ in the
higher redshift spike and $0.71\pm 0.22$ in the lower redshift
spike. Given the relatively large errors, these ratios are all
consistent with each other, even if the higher value in the lower
redshift spike suggests that X-ray activity may be enhanced in this
structure.

We note that most of the X-ray sources are identified as AGN on the
basis of their optical and X-ray properties. Whenever the optical
classification is uncertain, we considered as AGN those sources
satisfying at least one the following conditions: $L_x>10^{42}$ erg
s$^{-1}$, where $L_x$ is the observed luminosity in the 0.5-10 keV
band; $HR>0$, where HR is the X-ray hardness ratio; $f_x/f_r>0.1$,
where $f_x$ and $f_r$ are the fluxes in the 0.5-10 keV and R band,
respectively. Indeed, normal and starburst galaxies typically do not exceed
luminosities of $10^{42} \lum$ (e.g. Fabbiano et al. 1992) and have
X-ray to optical flux ratios an order of magnitude smaller than those
of AGN \citep{stock91}. Also, starburst spectra cannot be as hard as
those of heavily obscured AGN: we verified that an average starburst
template derived from \citet{dahl98} would always produce $HR<0$ in
our observations. Overall, 101 out of the 131 X-ray sources in our
sample are classified as AGN.
%22 are classified as galaxies, 7 are extended (group or clusters) 
%and 3 are stars. 
Fifteen AGN are found in the $z=0.67$ spike and 16 in the
$z=0.73$ spike, to be compared with the total number of 17 X-ray
point-like sources in each spike. As a consequence, the results of the
above analysis would be essentially the same if only sources
identified as AGN were considered.

%\begin{equation}
%F_X=4\pi f_{X,tot}\left({2 c \over H_0}\left(1+z-\sqrt{1+z}\right)\right)^2
%\approx
%1.6\times 10^{58}f_{X,tot}\left(1+z-\sqrt{1+z}\right)^2
%\end{equation}

\subsection{X-ray to K20 matched sample}

The results presented in the previous Section are based on the
comparison of data from two surveys (X-ray and K20) which cover
different areas.  In order to further check the possible significance
of the enhanced X-ray activity among the sources at z=0.67, we have
studied in more detail the correlation between the K20 sources and the 49
X-ray sources detected within the K20 field.

First we searched for any K20 counterpart of an X-ray source within a
radius of 10 arcsec, choosing as the most likely counterpart candidate
the K20 source closer to the X-ray source. 
%Since the X-ray source
%positions have been registered on the FORS images, they should have
%better position accuracy with respect to the K20 sources. 
Then we computed the RA and DEC offset histograms and fitted them with
a Gaussian. 
%We find a systematic shift between the X-ray and K20
%coordinates of $\Delta RA=RA_K-RA_X=-0.86$ arcsec and $\Delta
%DEC=DEC_K-DEC_X = 1.25$ arcsec. 
The Gaussian width resulted to be $\epsilon_{RA} = 0.55$ arcsec and
$\epsilon_{DEC} = 0.50$ arcsec for the RA and DEC histogram
respectively. We then corrected the K-band coordinates for the
systematic shift with respect to the X-ray coordinates and searched
for the K20 counterpart in a circle with radius = 1.77 arcsec
(3.4$\epsilon$, where $\epsilon$ is the average between
$\epsilon_{RA}$ and $\epsilon_{DEC}$) around the X-ray source
position. We recall that, given a circular Gaussian of width
$\epsilon$, a circle of radius $3.4\epsilon$ encloses 99.7\% of the
volume and then corresponds to the 3$\sigma$ error box.

The cross correlation produced 30 matches between X-ray and K-band
sources (then 19 X-ray sources in the K20 field have no
counterpart down to $K_s<20$). Given the surface density of K20 and
X-ray sources, we estimate that the expected number of spurious chance
coincidences among these X-ray and K20 matches is of the order of one.

We note that the positional match has been done using the coordinates
of the X-ray source centroids rather than those of the optical
counterparts, which are typically separated by $\sim 0.5$ arcsec from
the X-ray centroids. Using the coordinates of the optical counterparts
rather than those of the X-ray sources would not affect our results.

Although our adopted match radius might not be appropriate
in the case of extended X-ray sources, we note that only two X-ray
sources falling in the K20 field are extended. Both of them have K20
counterparts and are examined separately. With our adopted search
radius, a unique K20 counterpart is found for the extended source CDFS
566, which coincides with the cD galaxy at the center of the cluster
at $z=0.73$. The other extended source, CDFS 511, can be associated to
any of three K20 sources with redshifts between 0.765 and 1.047, and
its identification remains uncertain. 
%Since a concentration of K20 sources with $z\sim1.04$ is found just
%outside our search radius, we identify this extended X-ray source as a
%cluster at $z=1.04$. 
For all the 28 point-like X-ray sources with a K20
counterpart, this is unique. Only three out of 30 matches do not have
a high quality redshift determination either from the K-band or the
X-ray catalog. We verified that for the 20 sources with high quality
redshifts both in the X-ray and in the K20 catalog the two
measurements are always in excellent agreement, with essentially zero offset
and a dispersion of $\Delta z=0.003$.
It is interesting to note that a large fraction (8/18) of the sources
classified as AGN in the X-ray catalog were not recognized as such in
the K20 spectroscopic data alone.
%[COMMENT ON THE K20 AGN NOT DETECTED IN THE X-RAYS?] 
%Also, given the extremely deep exposure in the X-rays, we can sample
%rest frame luminosities in the 0.5-10 keV band down to $\sim 10^{42}
%\lum$ at $z\sim 1.3$, where the K20 sensitivity drops. We can
%therefore assume that essentially {\it all} the AGN dominated sources
%among the K20 ones are detected by \chandra.

We then estimated the fraction of K-band selected sources detected in
the X-rays inside and outside the two structures, excluding extended
X-ray sources. Again an excess of X-ray sources in the low redshift
spike appears. The fraction of X-ray detected sources in the low
redshift spike is $0.21\pm 0.10$ (5/24), while it is 
$0.08\pm 0.04$ both in the high redshift spike (4/47) and in the
neighboring redshift bins (i.e. in the range $0.4<z<1.0$).

Since we are now dealing with a sample free from selection biases, we
tried to estimate the significance of the X-ray source overdensity at
$z=0.67$. Since 25 X-ray to K20 matches are point-like
sources with measured redshift, we simulated 10000
samples of 25 sources randomly extracted from the K20 spectroscopic
catalog, finding that in only 5.7\% of the cases these random samples
contain 5 or more sources at z=0.67 (i.e. in the redshift interval
0.664--0.685). We can therefore conclude that the X-ray source 
enhancement in the low redshift spike is statistically significant 
at about $2\sigma$ level.
A similar level of significance is found by restricting this analysis
to the 18 matches identified with AGN (4 of which are at z=0.67).

\section{Discussion} \label{discussion}

Our results show that in the CDFS several large scale structures of
X-ray sources (mostly AGN) are being detected, which appear as narrow
spikes in the source redshift distribution. The most prominent
structures are detected at z=0.67 and z=0.73, while other peaks appear
at higher redshifts.\footnote{We note in passing that even removing
the two structures at z=0.67 and z=0.73, the X-ray source redshift
distribution in the CDFS shows an excess at $z<1$ with respect to the
predictions of standard synthesis models of the X-ray background
\citep{comas95, gilli01}.} In the CDFN \citet{barger02} found evidence
for similar redshift spikes at z=0.843 and 1.017 with at least ten
X-ray sources each. Despite several efforts in the past years, before
the deep X-ray surveys by \chandra there was no convincing evidence
for three-dimensional clustering of X-ray selected AGN. Only a
$2\sigma$ detection was found by \citet{carrera98} in the ROSAT
International X-ray Optical Survey (RIXOS, Mason et al. 2000) on
scales $<60-120\:h_{70}^{-1}$ Mpc. Interestingly, the $2\sigma$ signal
detected in the RIXOS refers to the subsample of sources in the
redshift range 0.5--1.0, where the biggest structures in the \chandra
deep fields are also detected, although the lack of clustering signal
at $z<0.5$ and $z>1$ might be due to the small volume sampled and to
the falling sensitivity of the RIXOS, respectively.

Significant clustering signal of X-ray selected AGN was found from two
dimensional analysis based on the angular correlation function
(e.g. Akylas et al. 2000; Vikhlinin \& Forman 1995). The clustering
length of local X-ray selected AGN was found to be similar to that of
local galaxies, supporting the idea that active nuclei are
unbiased with respect to normal galaxies. This was also reported by
\citet{smith95} who found that the amplitude of the low-redshift
($z<0.3$) QSO-galaxy angular cross-correlation function is identical
to that of the APM galaxy angular correlation function, implying that
QSO inhabit environments similar to those of normal
galaxies. \citet{brown01} showed that in the redshift range
$0.2<z<0.7$ the spatial cross-correlation of AGN with early type
galaxies is much stronger than that between AGN and late type
galaxies. We found that 80\% of the highly significant peaks seen in the
K20 source redshift distribution, in the range $0.5<z<1.3$, have a
corresponding peak in the X-rays. This supports the idea
that at these redshifts AGN and early type galaxies, whose detection
rate is higher in K-band than in optically selected samples, are
tracing the same underlying structures.

%The fact that on the
%average the fraction of X-ray to K20 sources is similar inside and
%outside the redshift peaks suggests that in general X-ray sources are
%not significantly biased with respect to normal galaxies. This is true
%also when considering only AGN among our X-ray selected sources, in
%agreement with the findings of \citet{akylas00}.

Since the large scale structures at $z=0.67$ and $z=0.73$ appear as
narrow spikes in the source redshift distribution and span the whole
\chandra field of view, we verified whether they can be actually
considered as ``sheets'', i.e. nearly bi-dimensional structures, or
filaments of galaxies as predicted by cosmological models of structure
formation. The structures at $z=0.67$ and $z=0.73$ have an observed
thickness of $18.8$ Mpc and $13.4$ Mpc respectively (the redshift
range covered by sources at $z=0.67$ is slightly larger than that
covered by sources at $z=0.73$), to be compared with their 7.3 Mpc
extension on the plane of the sky. Given the errors on the redshift
measurements, the values obtained for the physical thickness of the
structures are likely to be upper limits. Furthermore, if these
structures do not lay exactly on the plane of the sky, any projection
effects would broaden their apparent thickness. We verified that for
the broader structure at $z=0.67$ sources at higher right ascension
tend to have smaller redshifts. This projection effect needs however
better statistics to be quantified. We finally note that 7.3 Mpc is
likely to be a lower limit to the angular extension of the two
structures, which cover almost entirely the \chandra field of
view. Given the above uncertainties, an enlarged field and increased
source statistics are needed to determine the geometry of the two
structures.

We searched for enhanced X-ray activity at $z=0.67$ and $z=0.73$ by
cross-correlating the X-ray and the K20 catalogs and looking at the
fraction of K20 sources with X-ray counterparts at different
redshifts. Five out of 47 K20 sources at $z=0.73$ match an X-ray
source: one is the extended source CDFS 566 and is associated to the
cD galaxy of the rich cluster seen in the K20 survey; one is
classified as a normal galaxy and the remaining three sources are
identified as AGN. Five out of 24 K20 sources at $z=0.67$ do show
X-ray emission: one is classified as a normal galaxy, while the
remaining four are classified as AGN. We found a weak ($2\sigma$)
X-ray source overdensity in the structure at $0.67$ with respect to
the field and to the other structure at $z=0.73$. This overdensity is
also observed with the same significance when considering only those
X-ray sources identified as AGN.  While in the X-rays the two
structures are very similar (same number of sources, similar angular
extent and redshift thickness), in the K20 survey they appear very
different. The K20 sources at $z=0.73$ are dominated by a virialized
galaxy cluster, with a central cD galaxy and spectral segregation,
with early type galaxies clustered in the inner regions. On the
contrary the K20 sources at $z=0.67$ do not appear to constitute a
galaxy cluster, being uniformly distributed across the K20 field. We
note incidentally that $> 50\%$ of the K20 sources classified as early type
galaxies lay in the two structures at $z=0.67$ and $z=0.73$ and that a
concentration of early type galaxies at $0.67$ in the CDFS was also
reported by \citet{cwg01}.
%In
%this structure the AGN fraction is therefore 17\%, a factor of $\sim
%4$ higher than in the cluster at $z=0.73$ and than in the field
%(although the statistics is very low). Our simulations established
%this enhancement to be significant at $>94\%$ confidence level.
The difference between the two structures when observed by the K20
survey can be ascribed to the small region covered by the K20 survey,
which has sampled a rich cluster at z=0.73 without sampling any
cluster at $z=0.67$. Since the two structures are very similar in the
X-rays, we might speculate that, in a region free from clusters, we
would have observed an X-ray source overdensity even at $z=0.73$. In
principle, this suggests that X-ray activity in the large scale
structures is triggered preferentially away from the higher density
peaks corresponding to galaxy clusters. However, further work is
necessary to increase the source statistics and test this idea.
%
%This is also confirmed by the fact
%that in the cluster at $z=0.73$ no AGN (X-ray sources) are detected
%within 600 kpc from the cluster core marked by the cD galaxy (see
%Fig.~3). 
%

\section{Conclusions and future work} \label{conclusions}

Our main results can be summarized as follows:

\noindent
1) Several large scale structures of X-ray sources (mostly AGN) are
being detected in the \chandra Deep Field South, which appear as
narrow spikes in the source redshift distribution. The most prominent
structures are detected at z=0.67 and z=0.73. In addition, high
redshift structures are significantly detected at z=1.04, 1.62 and
2.57. This represents one of the first evidences for spatial
clustering of X-ray selected AGN.

\noindent
2) Similar redshift structures, the most prominent at $z=0.67$ and
$z=0.73$, are observed among the K-band selected galaxies of the K20
survey. About 80\% of the most significant peaks in the redshift
distribution of K20 sources have a corresponding peak in the
X-rays. Since only a fraction of the X-ray sources in these peaks is
within the smaller K20 field, we can conclude that, in general, the
structures seen in the K20 survey are traced on wider scales by the
X-ray sources. We also notice that, since the K20 survey sensitivity
drops above $z\sim 1.3$, high redshift structures are detected only in
the X-rays.

\noindent
3) By cross-correlating the K20 and the X-ray catalog we find a weak
evidence ($\sim 2\sigma$) of enhanced nuclear activity among the K20
sources at $z=0.67$ with respect to the field, suggesting a
preferential trigger in this structure. Given its low significance,
this result has to be confirmed by further observations.

Significant improvements in the understanding of the large
scale structures in the CDFS is expected from the on going and planned
multiwavelength observations of the field. The XMM deep pointing (500
ksec in a 30 arcmin diameter region; Hasinger et al. 2002) is actually
expanding the area covered by X-ray data by a factor of $\sim
2$. Accurate photometric redshifts for the optically faint X-ray
sources are being obtained from FORS optical images and ISAAC 
near-IR data and will be refined further using data from the 
Advanced Camera for Survey (ACS) on HST.

%Photometric redshifts will
%extend the comparison between the K-band and the X-ray source
%population outside the K20 field.

%We note in passing
%that the bigger X-ray field of view is crucial in determining the
%nature of the large scale structure we see, since they extend much
%farther than observed than observed in the K20 region. Therefore, any
%estimate of the cluster mass and luminosity from the K20 data would
%have been severely underestimated.
%Since the same redshift structures of the K20 survey have been
%detected in the X-ray but on a much wider area, it is likely that a
%complete K-band mapping of the CDFS would extend the K20 structures on
%much wider scales. 
%ADD NOTE : X-RAY SPIKE AT Z=0.73 IS NARROWER THAN THE CORRESPONDING K20 SPIKE
%WHILE AT Z=0.67 THE WIDTH IS SIMILAR

\acknowledgements

We wish to thank the referee for useful comments.

%R. Giacconi and C. Norman gratefully
%acknowledge support under NASA grant NAG-8-1527 and NAG-8-1133.

\begin{table}
\begin{tabular}{ccclcc}
\hline \hline
 & \multicolumn{2}{c}{K20~~~~~~~}&&\multicolumn{2}{c}{X-ray~~~}\\
~~~~z~~~~~~& ~N~~& ~Prob.~~~~~&& N& Prob.\\
\hline
0.077&	5&	$4.9 \;10^{-4}$&&	 3&	$1.0 \;10^{-2}$\\
0.218&	5&	$3.3 \;10^{-2}$&&	  &	       	       \\
0.367&	7&	$3.7 \;10^{-2}$&&	  &	               \\
0.524& 11&	$6.0 \;10^{-4}$&&	  &	               \\
0.670& 24&	$1.9 \;10^{-4}$&&	19&	$9.1 \;10^{-5}$\\
0.738& 47&	$6.0 \;10^{-8}$&&	19&	$1.7 \;10^{-6}$\\
1.036& 15&	$6.4 \;10^{-4}$&&	 6&	$4.2 \;10^{-3}$\\
1.220& 10&	$3.2 \;10^{-4}$&&	 4&	$2.2 \;10^{-2}$\\
1.618&	 &	               &&	 5&	$3.8 \;10^{-3}$\\
2.572&	 &	               &&	 4&	$9.7 \;10^{-3}$\cr
\hline
\end{tabular}
\caption{ Peaks detected in the X-ray and K20 source redshift
distributions, sorted by increasing redshift. The signal and
background distributions are smoothed with $\sigma_S=300$ km/s and
$\sigma_B=1.5 \: 10^4$ km/s, respectively. Together with the central
redshift of each peak, the number of sources N in each peak and the
Poissonian probability of observing N or more sources given the
background value are also shown.}
\end{table}

%%%%%%%%%%%%%%%%%%%%%%%%%%%%%%%%%%%%%%%%%%%%%%%%%%%
%Table version with the confidence level
%0.077&	5&	99.9502&	 3&	98.9823\\
%0.217&	5&	98.3980&	  &	       \\
%0.248&	 &	       &	 4&	98.8058\\
%0.367&	7&	99.3610&	  &	       \\
%0.524& 11&	99.9395&	  &	       \\
%0.670& 24&	99.9999&	19&	99.9999\\
%0.738& 47&	99.9999&	19&	99.9999\\
%1.036& 15&	99.9360&	 6&	99.5806\\
%1.220& 10&	99.9925&	 4&	98.7520\\
%1.618&	 &	       &	 5&	99.8683\\
%2.572&	 &	       &	 4&	99.6942\cr
%%%%%%%%%%%%%%%%%%%%%%%%%%%%%%%%%%%%%%%%%%%%%%%%%%%

\clearpage

\begin{figure}
\plotone{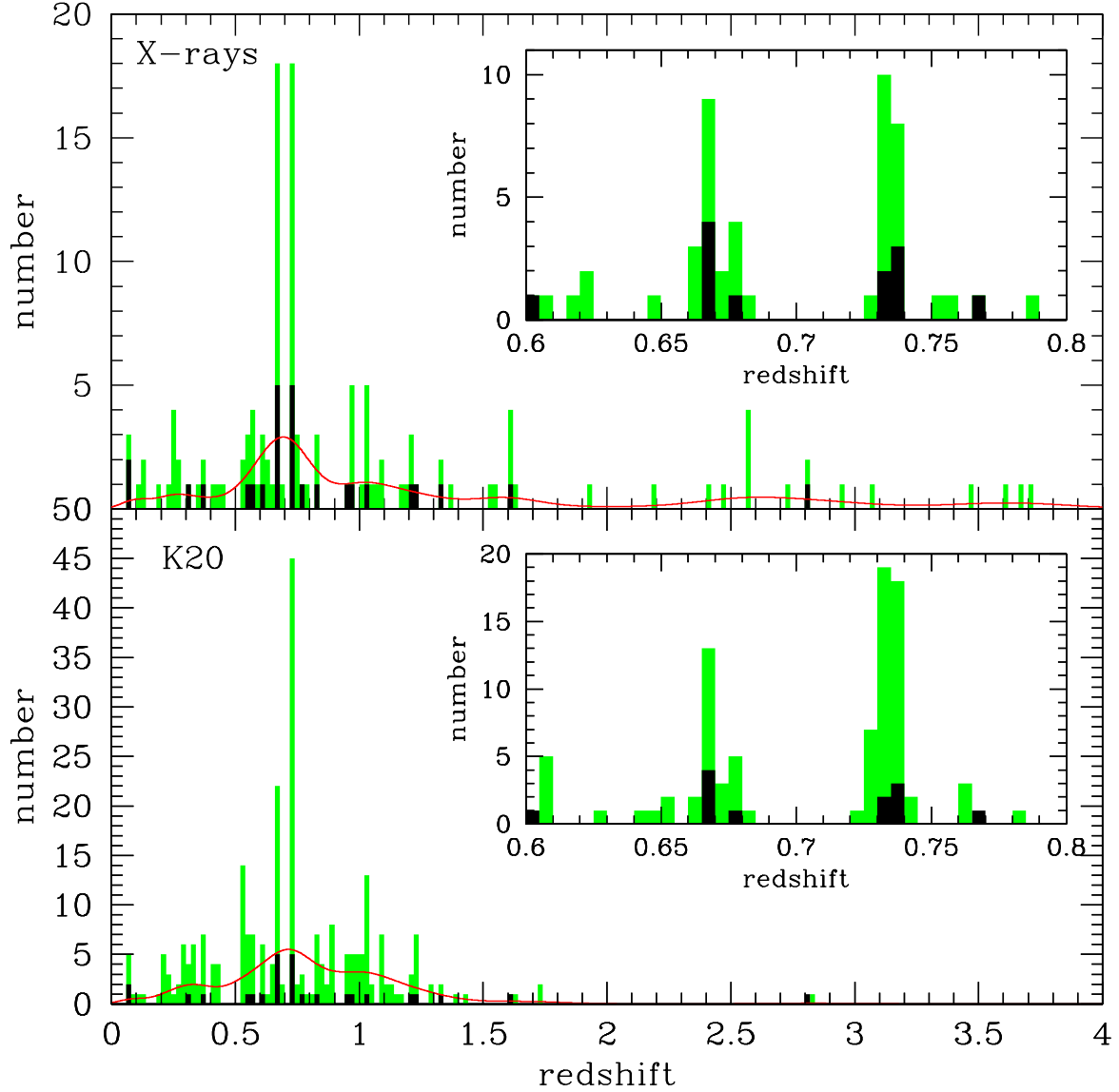}
\caption{Redshift distribution for X-ray sources (Upper Panel) and K20
sources (Lower Panel) with high quality optical spectra. The binning
shown in this Figure is $\Delta z=0.02$. The insets show a zoom on the
two main redshift spikes at z=0.67 and z=0.73 (binning is $\Delta
z=0.005$). The black histograms represent the matches between the
X-ray and the K20 catalogs. The ``background'' field distributions,
derived by smoothing the observed ones with
$\sigma_B=1.5 \: 10^4$ km/s (see text), are shown as continuous lines.}
\label{zdist_x}
\end{figure}

\begin{figure}
\plotone{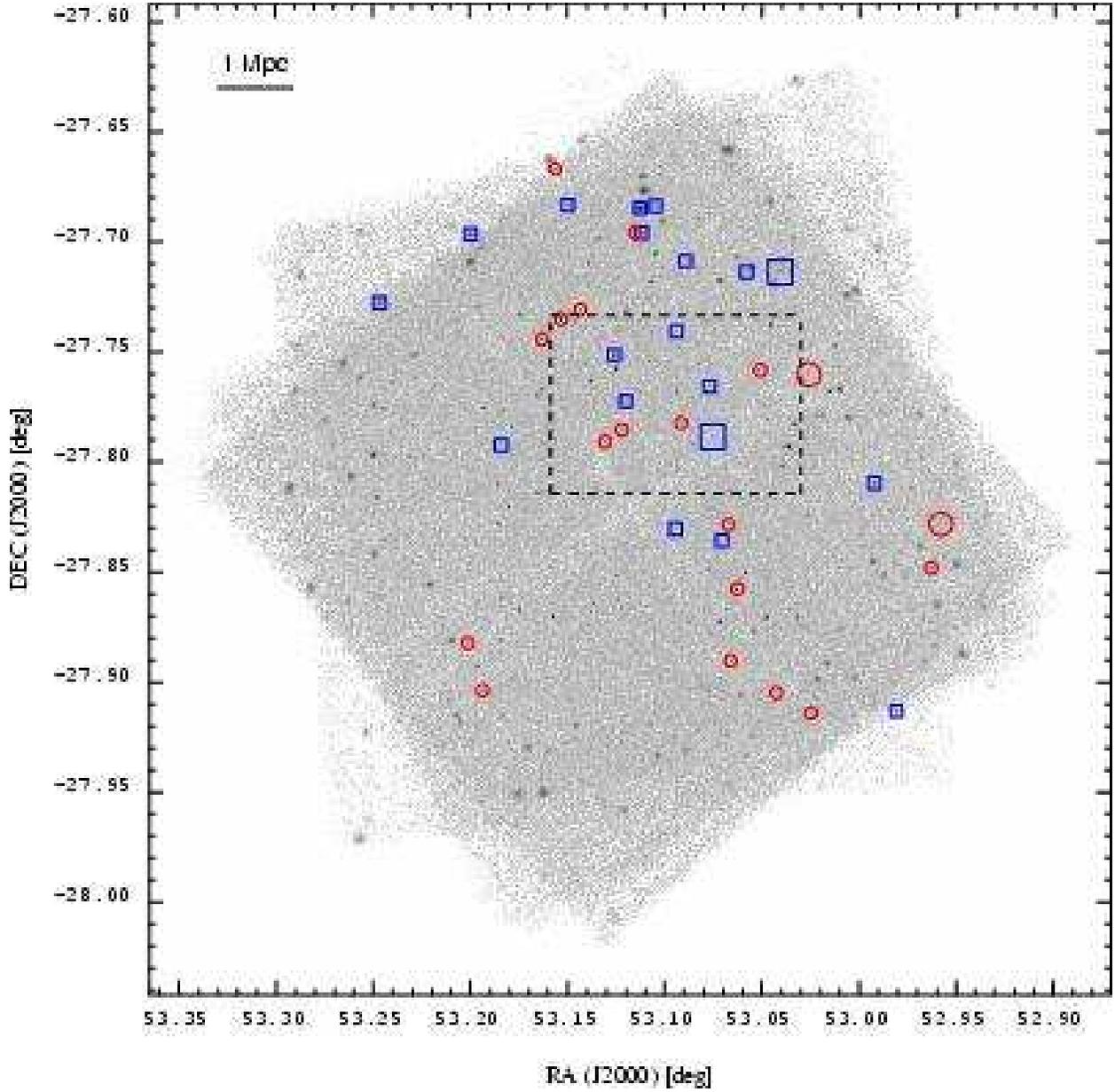}
\caption{\chandra ACIS-I image of the CDFS with the sources in the two
redshift spikes marked with different symbols: circles for sources at
$z=0.67$ and squares for sources at $z=0.73$. Extended sources in the
spikes are represented as big symbols. The ACIS-I detector is $\sim
17$ arcmin on a side. The dashed box indicates the $6.7 \times 4.8$
arcmin region covered by the K20 survey.}
\label{sdist_x}
\end{figure}

\begin{figure}
\plotone{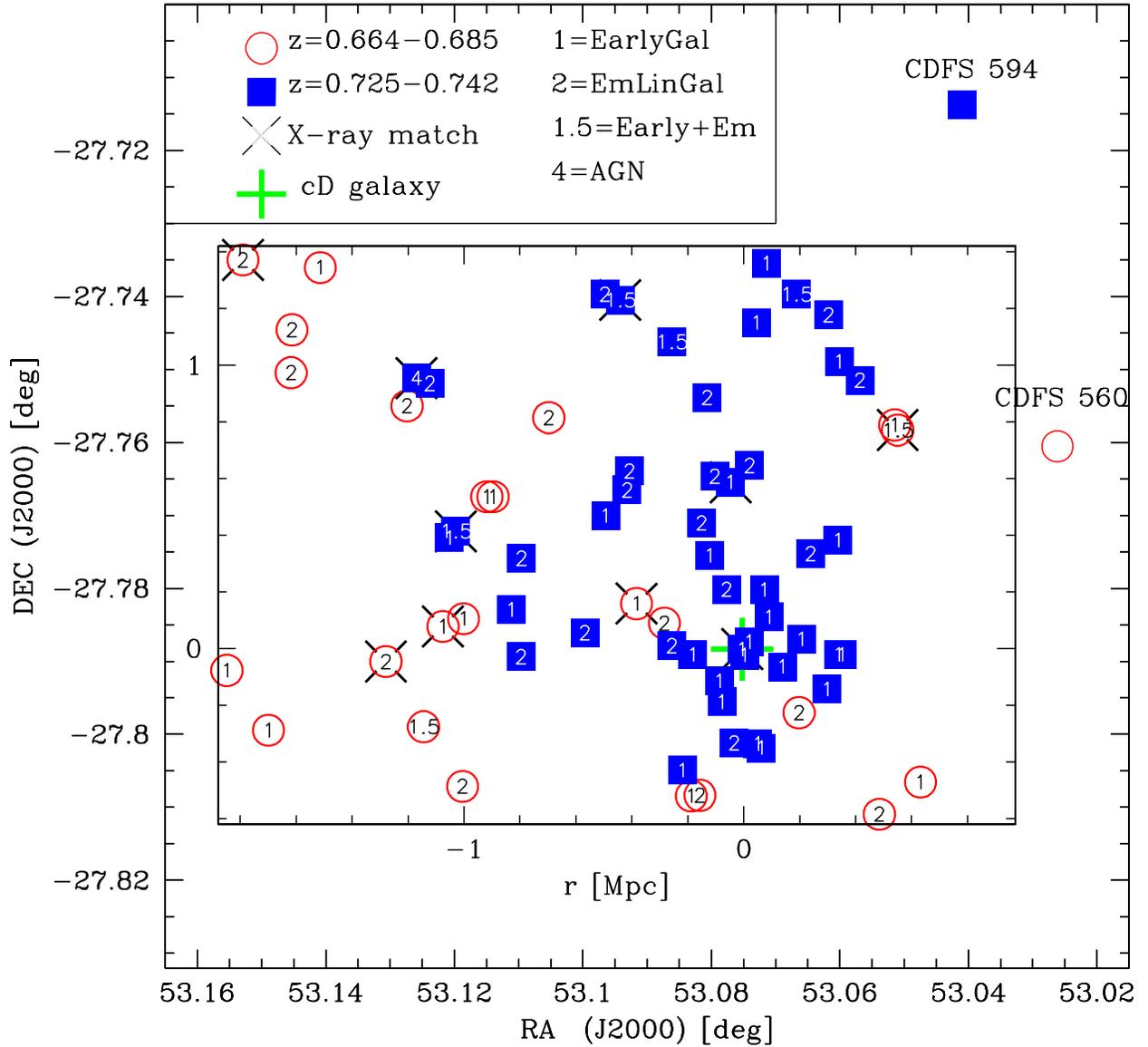}
\caption{Spatial distribution of the K20 sources at $z=0.67$ (open
circles) and at $z=0.73$ (filled squares). Their
classification from K20 spectroscopy is also indicated by the numbers
inside the symbols. Crosses are overplotted on sources with X-ray
counterparts. Sources at $=0.73$ appear to constitute a standard
galaxy cluster with a cD galaxy marking its center (indicated by the
big cross) and showing spectral segregation, with early type
galaxies concentrated in the inner regions.  On the contrary, sources
at $z=0.67$ do not cluster around any cD galaxy and are uniformly
distributed across the field. One galaxy group at z=0.67 (CDFS 560) and
one at z=0.73 (CDFS 594) detected in the X-rays just outside the K20
field are also indicated.}
\label{k20fig}
\end{figure}

\end{document}